\documentclass[useAMS,usenatbib]{mn2e}
\usepackage{hyperref}
\usepackage{graphicx}
\usepackage{times}
\usepackage{epsfig}
\usepackage{rotating}
\usepackage{sidecap}
\usepackage{longtable}
\usepackage{lscape}
\def\src{HETE J1900.1-2455}
\def\xmm{\it XMM-Newton}
\def\xte{\it RXTE}

\def\ltsima{$\; \buildrel < \over \sim \;$}
\def\simlt{\lower.5ex\hbox{\ltsima}}
\def\gtsima{$\; \buildrel > \over \sim \;$}
\def\simgt{\lower.5ex\hbox{\gtsima}}

\title[Accretion flow to AMSP, HETE J1900.1--2455]{The accretion flow
  to the intermittent accreting ms pulsar, HETE J1900.1--2455, as
  observed by {\xmm} and {\xte}} \author[A. Papitto et
  al.]{A.~Papitto$^{1}$\thanks{E-mail: papitto@ice.csic.es}, A. D'A\`i$^{2}$, T.~Di Salvo$^{2}$, E.~Egron$^{3}$, E.~Bozzo$^{4}$
  L.~Burderi$^{3}$, R.~Iaria$^{2}$, \newauthor A.~Riggio$^{3}$, M.~T.~Menna$^{5}$
  \\ $^{1}$ Institut de Ci\`encies de l'Espai (IEEC-CSIC) Campus UAB,
  Fac. de Ci\`encies, Torre C5, parell, 2a planta 08193 Barcelona,
  Spain\\ $^{2}$ Dipartimento di Fisica, Universit\`a di
  Palermo, via Archirafi 36, 90123 Palermo, Italy\\ $^{3}$ Dipartimento di Fisica, Universit\`a di
  Cagliari, SP Monserrato-Sestu, Km 0.7, 09042 Monserrato,
  Italy\\ $^{4}$ ISDC Science Data
  Center for Astrophysics of the University of Geneva, chemin
  d’\'Ecogia 16, 1290 Versoix, Switzerland\\ $^{5}$ INAF Osservatorio Astronomico di Roma, via Frascati
  33, 00040 Monteporzio Catone, Italy}
\begin{document}

\date{\today}

\pagerange{\pageref{firstpage}--\pageref{lastpage}} 

\maketitle

\label{firstpage}

\begin{abstract}

We present a study of the accretion flow to the intermittent accreting
millisecond pulsar, {\src}, based on observations performed
simultaneously by {\xmm} and {\xte}. The 0.33--50 keV energy spectrum
is described by the sum of a hard Comptonized component originated in
an optically thin $\tau\simeq1$ corona, a soft thermal $kT_{\rm
  in}\simeq0.2$ keV component interpreted as accretion disc emission,
and of disc reflection of the hard component. Two emission features
are detected at energies of 0.98(1) and 6.58(7) keV, respectively.
The latter is identified as K$\alpha$ transition of \mbox{Fe\,{\sc
    xxiii}--{\sc xxv}}. A simultaneous detection in EPIC-pn,
EPIC-MOS2, and RGS spectra favours an astrophysical origin also for
the former, which has an energy compatible with Fe-L$\alpha$ and
helium-like Ne-K$\alpha$ transitions. Broadness of the two features,
$\sigma/E\simeq0.1$, suggests a common origin, resulting from
reflection in an accretion disc with inclination of
$(30^{+4}_{-3})^{\circ}$, and extending down to $R_{\rm
  in}=25^{+16}_{-11}$ gravitational radii from the compact
object. However, the strength of the feature at lower energy measured
by EPIC-pn cannot be entirely reconciled with the amplitude of the Fe
K$\alpha$ line, hampering the possibility of describing it in terms of
a broad-band reflection model, and preventing a firm identification.
Pulsations at the known 377.3 Hz spin frequency could not be detected,
with an upper limit of 0.4 per cent at 3-$\sigma$ confidence level on
the pulsed fractional amplitude.

We interpret the value of the inner disc radius estimated from
spectral modelling and the lack of significant detection of
  coherent X-ray pulsations as an indication of a disc accretion flow
truncated by some mechanism connected to the overall evolution of the
accretion disc, rather than by the neutron star magnetic field. This
is compatible with the extremely close similarity of spectral and
temporal properties of this source with respect to other, non pulsing
atoll sources in the hard state.

\end{abstract}


\section{Introduction}

Discovery of millisecond coherent pulsations from a number of neutron
stars (NS) in low mass X-ray binaries \citep[LMXB, ][]{wijnands1998}
proved that accretion of disc angular momentum is an efficient
mechanism to spin up a NS to ms spin periods.  X-ray pulsations from
an accreting NS are observed when the NS magnetic field is strong
enough to control at least partly the mass flow in the NS
surroundings, channelling it towards the magnetic poles. So far,
however, coherent ms signals have been detected only from a subset of
LMXB; of the $\sim 60$ systems confirmed to host a NS \citep[][update
  RKcat7.18, 2012]{ritter2003}, coherent pulsations with a period in
the ms range were observed from only 14 sources, commonly known as
accreting millisecond pulsars \citep[AMSP; see][for a recent
  review]{patruno2012c}.  Whether a magnetosphere is able to control
mass flow around the NS depends on the balance between the torques
exerted on the accretion flow by the magnetic field and by disc matter
viscosity. Observation of ms pulsations, thermonuclear X-ray bursts,
and quasi periodical oscillations, strongly points toward the magnetic
field of most NS in LMXB to lie in the range $ 10^{8}-10^{10}$ G.
Lack of a detection of pulsations from majority of LMXB, with
present-day observatories, is thus most easily interpreted in terms of
a larger average mass accretion rate; this enhances diamagnetic
screening of NS magnetic field by the accretion flow, and burial of
the field under the NS surface \citep{cumming2001}.  AMSP are indeed
relatively faint X-ray transients. Most of them show few-weeks to
few-months long outbursts, separated by years-long intervals spent in
a quiescent state. Their peak outburst X-ray luminosity never exceeds
$\sim 10^{37}$ erg s$^{-1}$, and the long-term average mass accretion
rate is usually $\simlt 10^{-3}$ times the Eddington rate
\citep{galloway2006}.  However, lack of detected pulsations from very
faint accreting NS \citep[e.g.,][]{patruno2010} indicates how such a
picture is not entirely satisfactory; alternative scenarios proposed
to explain the paucity of ms pulsars in LMXBs point to pulse
decoherence induced by scattering in a hot cloud around the NS
\citep{titarchuk2002}, magneto-hydrodynamical instabilities
\citep{kulkarni2008,romanova2008}, alignment between the magnetic
field and spin axes of the NS \citep{lamb2009}, and a low pulse
amplitude caused by gravitational light bending
\citep{wood1988,ozel2009}.

A source of crucial importance to investigate differences between
pulsating and {\it ordinary} LMXBs is {\src}. Discovered in 2005 by
{\it High Energy Transient Explorer 2} ({\it HETE-2},
\citealt{suzuki2007}), {\it Rossi X-ray Timing Explorer} ({\xte})
detected pulsations at a frequency of 377.3 Hz
\citep{kaaret2006}. Subsequent observations showed how pulsed
fractional amplitude decreased from $\approx4.5$ per cent to below the
sensitivity level, $\simlt 1$ per cent. This source was then the first
AMSP to show intermittent pulsations \citep{galloway2007}, later
joined by Aql X--1 \citep{casella2008} and SAX J1748.9--2021
\citep{altamirano2008}.  The magnetic burial scenario is particularly
appealing to explain the behaviour observed from {\src}, which is the
only AMSP to have shown an outburst lasting more than seven years; in
fact, disappearance of pulsations after two months of accretion at a
rate of few per cent the Eddington rate is compatible with the
timescale of magnetic screening \citep{cumming2008}. After their first
disappearance a few months after the outburst onset, pulsations from
this source were sporadically observed to re-appear at a very low
fractional amplitude, $\approx 0.5$ per cent
\citep{galloway2008,patruno2012}. This suggests how the magnetic field
can temporarily re-emerge to channel accretion flow towards NS poles,
or that emission is always pulsed, but most of the time below
detectability threshold of current observations.

A powerful probe of the accretion flow close to a compact object is
the reflection component sometimes detected in the X-ray energy
spectra of these sources. Such component arises from partial
back-scattering and reprocessing of hard ionising X-rays illuminating
the disc \citep[see, e.g., the review by][]{fabian2010}. As the disc
matter moves in high-velocity Keplerian orbits in the gravitational
well of the compact object, the shape of reflected spectrum is
modified by Doppler and gravitational redshifts \citep{fabian1989},
and conveys crucial information about the flow geometry. In
particular, the most prominent reflection feature is an emission line
at the energy of K$\alpha$ transition of iron; such feature is mainly
due to fluorescence of mildly ionised to neutral iron ($\simeq 6.4$
keV), or to recombination of helium-like ions of iron \citep[6.67--6.7
  and 6.97 keV for \mbox{Fe\,{\sc xxv}} and {\sc xxvi}, respectively;
  see, e.g.][]{kallman2004}.  {\xmm} and {\it Suzaku} observations of
the AMSP, SAX J1808.4--3658, have shown how the broad shape of an iron
line could be used to probe the transition region between disc and
magnetospheric flow around a quickly rotating accreting pulsar
\citep{papitto2009, cackett2009, patruno2009, wilkinson2011}. In
particular, the inner disc radius of the reflecting layer was
estimated to lie between 12 and 26 km from an assumed 1.4 M$_{\odot}$
NS, compatible with the theoretical expectations for a NS with a few
$\times10^8$ G magnetic field, accreting at the rate indicated by the
observed X-ray flux.

Here, we present the analysis of an {\xmm} observation of {\src}
performed in September 2011, aimed at determining the properties of
the accretion flow to the NS on the basis of the shape of the disc
reflection component and of its time variability properties. To obtain
a coverage at higher energies, we also take advantage of a
simultaneous observation performed by {\xte}.

\section[]{Observations and data analysis}

\subsection{XMM-Newton}

{\xmm} \citep{jansen2001} observed {\src} for 65.3 ks starting on 2011
September 19, at 15:31 UTC (Obs.Id. 0671880101). Data were processed
with the latest version of the {\xmm} Science Analysis Software
released at the moment of performing the analysis (v.12.0.0).

\begin{figure}
\includegraphics[angle=0,width=\columnwidth]{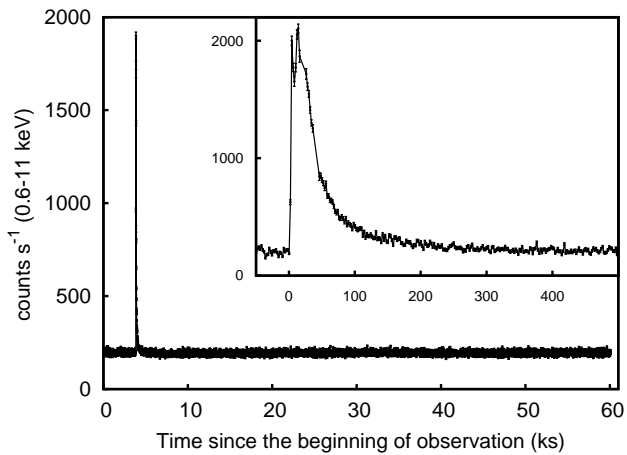}
\caption{0.6--11 keV light curve observed EPIC-pn. Inset shows the
  burst light curve with the time axis shown in seconds since the
  burst onset.}
\label{fig:lc}
\end{figure}

EPIC-pn camera \citep{struder2001} was operated in Timing Mode to
achieve a temporal resolution of $29.52$ $\umu$s.  We discarded last 5
ks of data from the analysis, since a marked increase of single
pattern events in the 10--12 keV band indicated contamination by
flaring particle background. The 0.6--11 keV light curve, corrected
with \texttt{epiclccorr}, is shown in Fig.~\ref{fig:lc}. A burst
clearly appears above an average count-rate of 192.5 s$^{-1}$ (see the
inset of Fig.~\ref{fig:lc}, where the burst light curve is plotted at
a magnified scale).  When analysing {\it persistent} emission, we
discarded 5 s before and 400 s after the burst onset.  In Timing
observing mode imaging information is preserved in one dimension only
(RAWX), while on the other (RAWY) data are collapsed into a single row
to allow a faster read-out.  To extract a spectrum of the source
emission we first considered 0.6--11 keV photons falling in a
rectangular region covering RAWX = 29 -- 46, discarding events
  flagged as bad (FLAG$=$0) and retaining only single and double
  pattern events (PATTERN$\leq$4). We used the task
\texttt{epatplot} to evaluate the observed fraction of single and
double events with respect to expectations. A deviation up to $2\%$
with respect to the model was found in the spectrum extracted from a
region covering also the brightest RAWX columns. We could obtain a
fraction of single and double events compatible with the model only by
excising the two brightest ones (RAWX=37--38); in order to avoid any
slight spectral distortion of the continuum due to photon pile-up we
considered the spectrum extracted from this reduced area. Spectra were
re-binned with the task \texttt{specgroup}, to not over-sample the
instrument energy resolution by more than a factor of three, and to
have at least 25 counts in each energy bin. The ancillary response
file (ARF) was produced by subtracting the ARF obtained from the
excluded region to the ARF calculated without excluding central
columns. To perform a temporal analysis we retained events with
  all patterns and falling also in the brighter RAWX columns of the
  CCD.

Of the two EPIC-MOS cameras \citep{turner2001}, MOS2 was operated in
timing mode to reduce pile-up, achieving a time resolution of 1.75
  ms, while MOS1 could not observe in such mode and was kept switched
off to allocate more telemetry bandwidth to the other CCDs. MOS2
observed an average count rate of 55.8 $s^{-1}$; we considered 
  single pattern (PATTERN$=$0) photons falling in RAWX = 276 -- 303
and 306 -- 335 to extract a spectrum, excising the central CCD columns
to limit pile-up, and removing bad events (FLAG$=$0). Background
was extracted from a box-shaped, 3000 x 2000 pixel wide region free of
sources, located in one of the outer CCD operated in imaging mode. We
used similar re-binning parameters than those applied to produce
EPIC-pn spectrum. We could not consider MOS2 data to perform a
  temporal analysis aimed at detecting a coherent signal, as the
  Nyquist frequency associated to its time resolution, 285.7 Hz, is
  lower than the spin frequency of the sorce (377.3 Hz).

Reflection grating spectrometer \citep{denherder2001} operated in
standard spectroscopy mode. We considered only first-order spectra,
re-binned in order to have at least 25 counts per channel. The average
persistent count rate was 3.9 and 4.8 c s$^{-1}$ for RGS1 and RGS2,
respectively.

\subsection{Rossi X-ray Timing Explorer}

Simultaneously to {\xmm} observation, {\src} was also observed for 3
ks by Rossi X-ray Timing Explorer ({\xte}), starting on 20 September
at 08:50 (ObsId.:96030-01-34-00). We extracted a spectrum from
Proportional Counter Array data \citep[PCA][]{jahoda2006}, considering
data taken by the top layer of PCU2 only, and producing a response
matrix with the latest calibration issued (Shaposhnikov et
al. 2009\footnote{
  \url{http://heasarc.gsfc.nasa.gov/docs/xte/pca/doc}\newline\url{/rmf/pcarmf-11.7/}}). Following
their recommendations we added a systematic error of 0.5$\%$ to each
spectral bin. A spectrum from data taken by High Energy X-ray Timing
Experiment \citep[HEXTE][]{rothschild1998} was extracted considering
cluster A, as it was the only in a on-source position during the
observation. Data observed by cluster B, which was instead permanently
in a off-source position, were used to estimate background.

\section{X-ray spectrum of persistent emission}

\subsection{The EPIC-pn 0.6--11 keV spectrum}
\label{sec:pn}
The X-ray spectrum of AMSP is dominated by Comptonization of soft
($kT_{soft}\simeq0.5$--$1$ keV) photons in a hot ($kT_e\simgt20$--$40$
keV), optically thin ($\tau\sim 1$) medium \citep{poutanen2006}. When
observations at energies below $\sim2$ keV are available, one or two
thermal components are detected and commonly interpreted as
originating from the accretion disc and/or the NS surface
\citep{gierlinski2002,gierlinski2005,papitto2009,patruno2009,papitto2010}. A
broad iron K$\alpha$ emission feature is often detected as well
\citep{papitto2009,cackett2009,cackett2010,papitto2010}, and together
with a Compton hump peaking between 20 and 40 keV
\citep{ibragimov2009,papitto2010,kajava2011,ibragimov2011} is
generally interpreted as a fingerprint of disc reflection of the hard
X-ray emission generated close to the compact object.

\begin{figure}
\includegraphics[angle=0,width=\columnwidth]{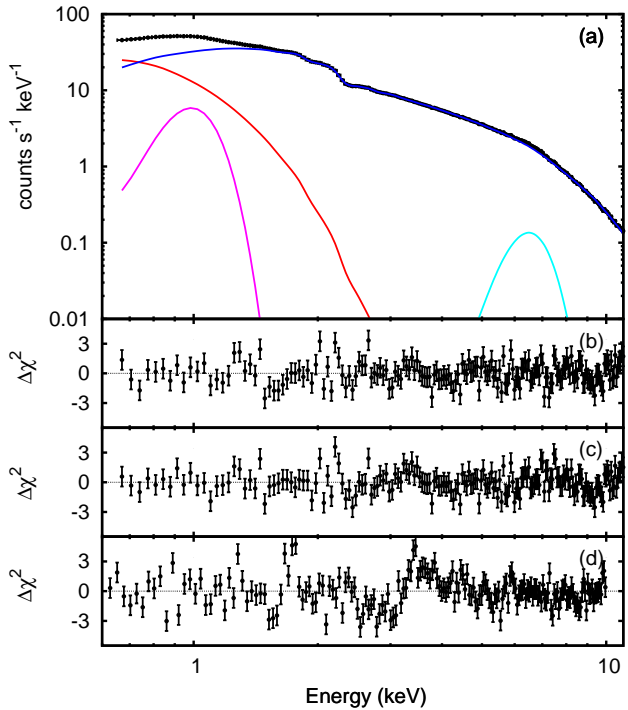}
\caption{0.6--11 keV spectrum observed by  EPIC-pn during the
  persistent emission. The Comptonized component (blue), the accretion
  disc multi-temperature black-body (red) and the Gaussian features at
  0.98 and 6.58 keV (magenta and light blue, respectively), used to
  model the spectrum according to Model A (see Table~\ref{tab1}) are
  also plotted (panel a).  Residuals of EPIC-pn data with respect to
  the best fitting Model A and B (see text and caption of
  Table~\ref{tab1}) are plotted in panel (b) and (c),
  respectively. Residuals of the spectrum simultaneously observed by
  EPIC-MOS2 with respect to Model A (to which a couple of Gaussian
  absorption feature at $\sim2$ keV were added) are plotted in panel
  (d).}
\label{fig:sp}
\end{figure}

\begin{figure}
\includegraphics[angle=0,width=\columnwidth]{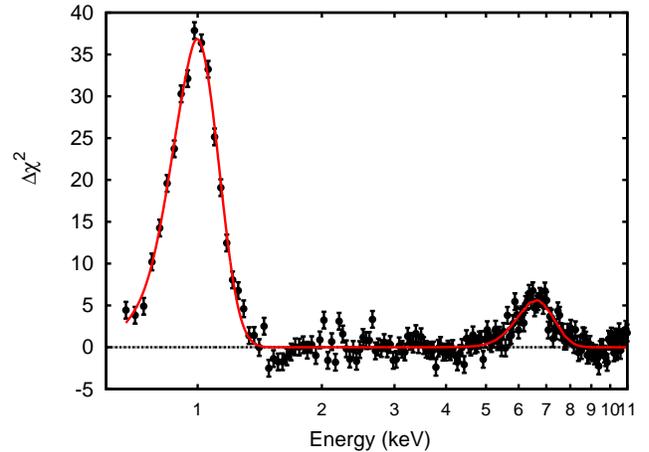}
\caption{Residuals of the 0.6--11 keV EPIC-pn spectrum with respect to
  Model A (see text and Table~\ref{tab1}),  when the Gaussian
  emission features centred at 0.98 and 6.58 keV are removed from the
  model.}
\label{fig:res}
\end{figure}

We first modelled the 0.6--11 keV continuum spectrum observed by
EPIC-pn (see the top panel of Fig.~\ref{fig:sp}) using a thermal
Comptonized component and a black-body, and describing photoelectric
absorption using an improved version of the T\"{u}bingen-Boulder model
(\texttt{TbNew} in XSpec terminology, Wilms, Juett, Schulz, Nowak,
2011, in preparation\footnote{
  \url{http://pulsar.sternwarte.uni-erlangen.de/wilms/}\newline\url{research/tbabs/}}). Abundances
and photoelectric cross-sections were set following \citet{wilms2000}
and \citet{verner1996}, respectively.  We modelled Comptonization with
the \texttt{nthcomp} model \citep[][see blue line in panel (a) of
  Fig.~\ref{fig:sp}]{zdziarski1996,zycki1999}, which describes the
up-scattered spectrum in terms of a power-law of index
$\Gamma\simeq2$, between a low-energy rollover at the seed black-body
photon temperature, $kT_{\rm soft}\simeq0.45$ keV, and a high-energy
cut-off at the electron temperature of the Comptonizing medium; since
this is left unconstrained even by data obtained by RXTE at higher
energies (see Sec.~\ref{sec:xmmxte}), we held it fixed to $kT_e=50$
keV. The thermal soft component has a temperature of $\simeq0.2$ keV
and a normalisation corresponding to a radius of $\simgt 30\:d_5$ km,
where $d_5$ is the distance to the source in units of 5 kpc
(\citealt{kawai2005}; see also \citealt{galloway2008} who derived a
compatible estimate of $4.7\pm0.6$ kpc). Since this size exceeds the
radius of a NS, and considering how the temperature is typical of
accretion disks around this kind of sources \citep[see,
  e.g.,][]{papitto2009}, we replaced the black-body component with a
multicolour accretion disc model (\texttt{diskbb} in XSpec; see red
line in panel (a) of Fig.~\ref{fig:sp}). Evident residuals at $\sim37$
and $\sim5$ $\sigma$ above the continuum model appear around
$E_1=0.98(1)$ and $E_2=6.58(7)$ keV. We modelled them with Gaussian
emission lines with a width of $\sigma_1=0.12(2)$ and
$\sigma_2=0.7(1)$ keV, respectively (see Fig.~\ref{fig:res}, showing
the residuals obtained when these two lines are removed from the
model). The description of the spectrum is also improved by an
absorption line centred at $E_3=1.79(2)$ keV, probably due to a
mis-calibration of the instrumental Si edge (see below).  The
chi-squared we obtained after the addition of these features is 1.40
over 169 degrees of freedom; residuals are shown in panel (b) of
Fig.~\ref{fig:sp}, and the values of the parameters obtained are
listed in the column labelled as {Model A} in Table~\ref{tab1}.

Given the presence of small residuals at low $\simlt 2$ keV energies,
and of a weak absorption feature at an energy close to the
instrumental Si edge (1.84 keV), we also modelled the spectrum
obtained after correcting the data with the task \texttt{epfast}. This
tool is designed to reduce the possible effect of charge transfer
inefficiency (CTI), and/or X-ray loading (XRL), on the spectra
observed by the EPIC-pn in fast readout modes. It was calibrated to
reduce residuals with respect to phenomenological models in the 1.5--3
keV, where the largest deviations are usually found (see the {\xmm}
calibration technical note, XMM-SOC-CAL-TN-0083, available at
\url{http://xmm.esac.esa.int/}). However, the corrected spectrum
resulted slightly more noisy than the unmodified one, and a further
emission feature centred at 2.23(1) keV, close to the instrumental Au
edge, should be added to the model; still the reduced chi squared of
the fit (1.44 over 167 d.o.f.) was larger than the one obtained
without correction, and for this reason we considered only the
spectrum obtained without \texttt{epfast} correction. We note how the
width and the normalisation of the feature at higher energy is left
unchanged by \texttt{epfast}, while its centroid energy slightly
increases with respect to the unmodified data, 6.66(7) keV. An
increase of the iron line energy in \texttt{epfast} corrected data was
already noted by \citet{walton2012}, who compared {\xmm} and {\it
  BeppoSAX} spectra of XTE J1650--500, concluding how the unmodified
EPIC-pn data possibly provided a better approximation of the true
energy scale.

Reflection of the hard X-ray continuum above the accretion disc is the
most convincing interpretation of the broad feature appearing at
6.58(7) keV, an energy compatible with K$\alpha$ transition of
iron. We used the model \texttt{reflionx} \citep{ross2005} to describe
reflection arising from an optically thick ionised slab, illuminated
by a $\Gamma=2$ power law spectrum. We convolved the reflection
component with an exponential low-energy cutoff (\texttt{expabs} in
XSpec), at an energy equal to the temperature of the soft seed photons
of the \texttt{nthcomp} component, k$T_{\rm soft}$, to take into
account the low energy rollover of the illuminating Comptonized
spectrum.  The reduced chi-squared obtained adding the reflection
component to Model A, and removing the Gaussian feature at the energy
of the iron K$\alpha$ transition, is 1.45 (250 over 172 d.o.f.). The
ionisation parameter lies in the range $\log{\xi}=$2.7--2.8,
compatible with the observed centroid energy of the iron line
\citep[see Table 1 of][]{garcia2011}. The addition of a Gaussian
emission line centred at $\approx 1$ keV was still requested by the
data, as the reflection model could not describe self-consistently the
soft excess; removing the line from the model yielded in fact to a
reduced chi-squared of $\simeq4.5$, due to evident residuals at those
energies. At the same time, the introduction of the reflection
component made the soft thermal component not significantly detected
anymore. However, the EPIC-pn response at energies $\simlt2$ keV is
clearly dominated by the Gaussian-shaped soft excess, found at a lower
energy (0.78(5) keV) and with a larger equivalent width (153 eV) than
the one obtained with model A, in which disc reflection was not
self-consistently modelled. The non detection of the soft thermal
component is then most probably due to correlation with the emission
feature, whose energy is evaluated by this model to lie very close to
the low energy end of the EPIC-pn energy band (0.6 keV). This is also
confirmed by the detection of the disc thermal component in the
spectrum obtained adding the RGS data-set, even when a reflection
component is included in the model (see Sec.~\ref{sec:xmmxte}).

\begin{table*}

  \caption{ Best fitting parameters of the X-ray spectrum of
    {\src}. Columns labelled as Model A (EPn),
    \texttt{TbNew(diskbb+nthcomp+gau$_1$+gau$_2$+gau$_3$)}, and Model
    B (EPn),
    \texttt{TbNew(nthcomp+expabs*rdblur*reflionx+gau$_2$+gau$_3$)},
    respectively, refer to the 0.6--11 keV spectrum observed by the
    EPIC-pn.  Column labelled as Model A (RGS) refers to the modelling
    of the RGS1/2 spectra in the 0.33--2.07 keV. Column labelled as
    Model B (XMM-XTE),
    \texttt{TbAbs*edge(nthcomp+expabs*rdblur*reflionx+gau$_2$+
      gau$_3$+gau$_4$)}, reports the parameters obtained by modelling
    of the spectra obtained simultaneously by RGS1/2 (0.33--2.07 keV),
    EPIC-pn (0.6--11 keV), PCA (3--30 keV) and HEXTE (20--50
    keV). Quoted fluxes are unabsorbed and evaluated in the 0.5--10
    keV band for Model A (EPn) and B (EPn), in the 0.5--2 keV for
    Model A (RGS), and in the 0.5--50 keV band for Model B
    (XMM-XTE). The inner disc radius evaluated from the normalisation
    of the \texttt{diskbb} component, $R_{\rm disc}$, is given for an
    inclination of $30^{\circ}$.  Uncertainties are evaluated at a
    $90$ per cent confidence level. Parameters held fixed in the fitting are
    reported without an uncertainty\label{tab1}} \resizebox{15cm}{!}{
\begin{tabular}{lrrrr}
 & Model A (EPn) & Model B (EPn) & Model A (RGS) &Model B (XMM-XTE)\\
\hline \\
$N_{\rm H}$ ($10^{21}$ cm$^{2}$)   & 1.3$_{-0.6}^{+0.4}$ & 0.6(2) & 1.02(5) & 1.05(3) \\ 
$A_{\rm O}$ & 1 & 1 &$2.11\pm0.15$ & $2.27\pm0.11$ \vspace{0.2cm}\\
$kT_{\rm in}$ (keV) & 0.21(1)& -- & 0.23$_{-0.02}^{+0.03}$ & 0.202(7)\\ 
$R_{\rm disc}$ (d$_5$ km) & 31.1$_{-8.5}^{+6.5}$ & -- & 29.5$_{-5.4}^{+6.7}$ & 33.4$_{-2.2}^{+2.8}$ \vspace{0.2cm} \\
$\Gamma$ & 1.99(1) & 1.979(7) & 1.99 & 1.975(4)\\
$kT_{\rm soft}$ & 0.45(1) & 0.458(4) & 0.51$_{-0.04}^{+0.06}$&0.446(5)  \\
$kT_{\rm e}$ & 50 & 50 & 50 & 50 \\
$N_C$ (10$^{-2}$ cm$^{-2}\:s^{-1}$) & 6.7(2)& 6.0(1) & 6.51$_{-0.6}^{+0.5}$ & 6.32(1)\vspace{0.2cm}\\

$E_1$ (keV) & 0.98(1) & 0.78$_{-0.06}^{+0.04}$ & 0.99(1)& 0.94(1) \\
$\sigma_1$ (keV) & 0.12(1) & 0.24(3) & 0.12(1)& 0.13(1) \\
$N_1$ ($10^{-3}\:cm^{-2}\:s^{-1}$) & 4.1$\pm$0.8 & 14$_{-3}^{+5}$& 3.8$_{-0.8}^{+1.0}$& 4.3(5)\\
EW$_1$ (eV) & 42 & 153& 38 & 40\vspace{0.2cm}\\
$E_2$ (keV) & 6.58(7) & {...}  & {...}  & {...} \\
$\sigma_2$ (keV) & 0.7(1) & {...}  & {...}  &{...}  \\
$N_2$ ($10^{-3}\:cm^{-2}\:s^{-1}$) & 0.6(1) & {...} &{...}  &{...}   \\
EW$_2$ (eV) & 135 & {...} & {...}  &{...}  \vspace{0.2cm}\\

$R_{\rm in}$ ($GM/c^2$) & {...} & 23$_{-7}^{+10}$ & {...}  &$25_{-11}^{+16}$ \\
$R_{\rm out}$ ($GM/c^2$) & {...} & 1000 & {...}  &1000\\
$\beta$ & {...} & -3.8$_{-5.2}^{+0.9}$ &{...}  & -3.7$_{-2.5}^{+1.1}$ \\
i ($^{\circ}$) & {...} & 27$\pm$3 & {...}  &30.6$_{-2.6}^{+3.7}$\vspace{0.2cm} \\
$\log(\xi)$ & {...} & 2.76$_{-0.04}^{+0.06}$ & {...}  &2.87$_{-0.08}^{+0.16}$)\\
$R=F_{\rm refl}/F_{\rm nthc}$ & {...} & 0.10(1) & {...}  &0.086(7) \vspace{0.2cm}\\
$E_3$ (keV) & 1.79(2) & 1.82(2) &{...}  & 1.80(2) \\
$\sigma_3$ (keV) & 0 & 0 &{...}  & 0\\
$N_3$ ($10^{-4}\:cm^{-2}\:s^{-1}$) & -1.1(3) & -1.2(3) &{...}  & -1.2(1) \vspace{0.2cm}\\

\hline
$F$ ($10^{-10}$ erg cm$^{-2}$ s$^{-1}$) & 7.3(2) & 6.97(8) & 2.46(2) & 13.59(8) \\ 
$\chi^2_r$ (d.o.f.)  & 1.40(169) & 1.36(169) & 1.16(4201) & 1.25(4460) \\

\end{tabular}
}
\end{table*}

To investigate if the broadness of the emission features detected in
the spectrum ($\sigma/E\approx0.1$) could be explained in terms of
relativistic effects expected to develop in the inner parts of the
accretion disc, we convolved the reflection model with the
\texttt{rdblur} kernel \citep{fabian1989}. This model describes the
relativistic effects due to the motion of plasma in a Keplerian
accretion disc immersed in the gravitational well of the compact
object, in terms of the inner and the outer radius of the disc,
$R_{\rm in}$ and $R_{\rm out}$ (in units of the gravitational radius,
$R_{\rm g}=GM/c^2$, where $M$ is the mass of the compact object), of
the index of the assumed power-law dependence of the disc emissivity
on the distance from the NS, $\beta$, and of the system inclination,
$i$, respectively, by assuming the Schwarzschild metric. We held the
value of the outer radius of the disc fixed to 1000 $R_{\rm g}$ as
spectral fitting could not constrain it. Convolving the reflection
model with such a component, the chi-squared of the fit decreased by
$\Delta\chi^2=-20.9$, for the addition of three free parameters. In
order to check the significance of such an improvement we used the
method of posterior predictive $p$-values \citep{protassov2002}. We
created 1000 fake spectra using the best fit model without disc
smearing, and fitted them with a model both including and excluding
the blurring component.  The probability that the chi-squared decrease
observed in real data after introducing disk blurring is due to
chance, is equal to the ratio between the number of cases in which a
chi-squared decrease equal or larger than the one observed in real
data was observed in fake spectra, and the number of trials. A
chi-squared decrease equal or larger than that observed in real
spectra was never obtained. We then estimated the probability that the
improvement in spectrum description obtained by adding the disc
smearing component to the model were due to chance as less than
$10^{-3}$, concluding how this component was significant at a
confidence level larger than 3-$\sigma$.The best-fitting parameters we
obtained with this model, labelled as B, are given in the second
column of Table~\ref{tab1}, and the residuals with respect to the
model are shown in panel (c) of Fig.~\ref{fig:sp}.

\subsection{Soft excess, MOS2 and RGS spectra}
\label{sec:soft}

In the previous section, we described phenomenologically the strong
excess appearing at $\approx 1$ keV in the EPIC-pn spectrum, by using
a broad Gaussian emission line. Alternative modelling in terms of a
bremsstrahlung continuum (\texttt{bremss} in XSpec) and an emission
spectrum from hot diffuse gas (\texttt{mekal}), did not gave
successful results in describing the excess, yielding a reduced
chi-square of $\chi^2_r=5.7$ and 3.0, over 170 d.o.f., respectively.
Several authors reported a similar excess at low energies in
observations performed by the EPIC-pn in fast modes \citep[see,
  e.g.,][]{hiemstra2011,walton2012}. The amplitude of this excess
seems to be correlated with the column density of the interstellar
absorption towards the considered source, as it is expected in the
case of an insufficient redistribution calibration (Guainazzi et
al. 2012, XMM-SOC-CAL-TN-083\footnote{available at
  \url{http://xmm.vilspa.esa.es/docs/documents}}).  However, the
column density of the interstellar absorption towards {\src} ($\simeq
10^{21}$ cm$^{-2}$) is lower than that affecting sources for which no
significant excess was found (e.g., SAX J2103.5+4545, which has a
$N_{\rm H}\simeq6\times10^{21}$ cm$^{-2}$), and thus no calibration
induced excess should be expected if a correlation with the column
density holds. Therefore, despite uncertainties in the instrument
calibration may be responsible for at least part of this excess, we
also explored plausible physical interpretations.

The centroid energy obtained with model A, 0.96(1) keV, is consistent
with K$\alpha$ emission of \mbox{Ne\,{\sc ix}--{\sc x}} (0.92--1.02
keV), and the \mbox{Fe-L$\alpha$} complex, which for $\log{\xi}$
ranging between 2 and 3 is made by transitions with energy between 0.8
and 1.2 keV \citep{kallman1995}. Such transitions are produced by disc
reflection, and are already included in the model \texttt{reflionx};
in particular the prominence of the \mbox{Fe-L$\alpha$} line is
expected to grow when the reflecting slab is iron-rich
\citep{ross2005,fabian2009}. Removing the Gaussian feature and letting
the iron abundance of the reflection component of Model B free to
vary, we obtained a best-fit abundance of 2.00(5) with respect to
Solar values, while other parameters of the reflecting medium were
compatible with those listed in Table~\ref{tab1} ($\beta\leq4.6$,
$R_{\rm in}=27^{+3}_{-5}\:R_{\rm g}$, $i=29(1)^{\circ}$,
$\log{\xi}=2.70(2)$, $R=F_{\rm refl}/F_{\rm nthc}=0.10(1)$; the
reflection fraction $R$ is evaluated as the ratio between the
unabsorbed fluxes in the reflected and Comptonized component, in the
0.5--10 keV band). However, the reduced chi-square of the fit, 2.34
over 169 degrees of freedom, is much larger with respect to Model B as
residuals appeared at energies below 2 keV even after the addition of
a \texttt{diskbb} component to the model. We also tried to fit the
spectrum obtained discarding energies above 5 keV; a successful
modelling is found with similar parameters than before, but this time
with a larger reflection fraction, $R=0.27(2)$. The large flux in the
$\approx1$ keV feature with respect to the iron line is then the most
probable reason for the non-achievement of a simultaneous modelling of
the hard and soft part of EPIC-pn spectrum in terms of the same
reflecting environment. At the same time, we could not model
successfully the whole EPIC-pn energy band even by using two
reflection components.

We then explored if part of the excess could be interpreted to an
overabundance of Ne in the same reflector originating the Fe K$\alpha$
line. Unfortunately, abundances of metals other than \mbox{Fe} cannot
be adjusted in \texttt{reflionx}. We then added a line blurred by the
same \texttt{rdblur} kernel used to convolve the reflection
component, to the previous model. A line centred at 0.95(1) keV
improved the modelling, but the chi-squared we obtained, 1.45 over 167
d.o.f., was still larger than that obtained simply modelling the
feature with a Gaussian (1.36 over 169 d.o.f., see Model B listed in
Table~\ref{tab1}), indicating how such a modelling was not completely
satisfactory.

In order to verify the origin of such an excess, we considered spectra
observed by other instruments on-board {\xmm}. The MOS2 spectrum is
much more noisy with respect to EPIC-pn, probably because of a more
uncertain energy calibration when operated in timing mode; two
additional absorption features were added in the \mbox{Au} K-edge
region around 2 keV, but still the final reduced chi-square was far
from being acceptable (2.62 over 166 d.o.f.; see bottom panel of
Fig.~\ref{fig:sp} for residuals with respect to Model
A). Nevertheless, we detected a Gaussian-shaped feature with
parameters entirely compatible with those found by EPIC-pn
($E_1=0.97(1)$ keV, $\sigma_1=0.12(2)$ keV, $N_1=4.7(2)\times10^{-3}$
cm$^{-2}$ s$^{-1}$, EW$_1=45(1)$ eV)

\begin{figure}
\includegraphics[angle=0,width=\columnwidth]{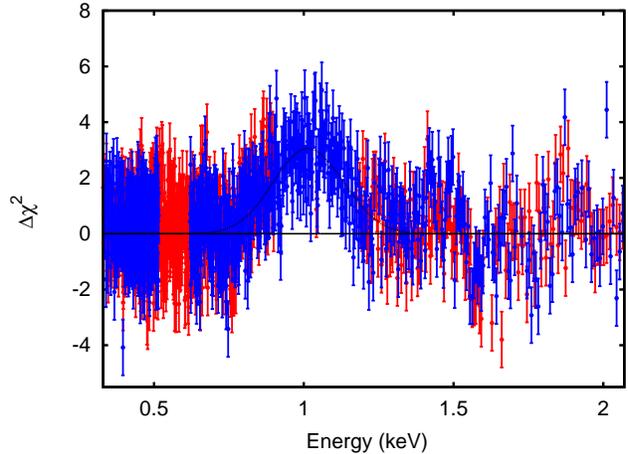}
\caption{Residuals of the 0.33--2.07 keV spectra observed by the RGS1
  (red points) and RGS2 (blue points) with respect to Model A (RGS;
  see caption of Table~\ref{tab1} and Sec.~\ref{sec:soft}), when the
  Gaussian emission feature centred at 0.99 keV is removed from the
  model.}
\label{fig:rgs}
\end{figure}

To model RGS spectra, we considered only first order spectra in
0.33--2.07 keV band, since the second order added no information.  We
modelled the continuum as in model A, with a multicoloured disc
black-body and a Comptonized component; given the limited bandwidth
covered by the RGS, we fixed the value of the asymptotic power law
index of the Comptonized component to the value indicated by the
EPIC-pn analysis (see leftmost column of Table~\ref{tab1}). We
modelled interstellar absorption with \texttt{TbNew}, and let a
relative normalisation constant between the two RGS spectra free to
vary. As residuals appeared around the \mbox{O\,{\sc i}} K edge at
(0.54 keV), and close to the \mbox{O\,{\sc i}} 1s--2p transition
(0.5275 keV), we let the abundance of O as a free parameter, obtaining
a best-fit value of $A_{\rm O}=2.11\pm0.15$ with respect to a H column
density of $N_{\rm H}=1.0(5)\times10^{21}$ cm$^2$. We found an excess
around $\sim 1$ keV in RGS data as well (see Fig.~\ref{fig:rgs}, where
the residuals obtained removing the lines from the best-fitting
models, are shown); by modelling it with a Gaussian feature,
parameters entirely compatible with those obtained modelling the
EPIC-pn data with model A, were found. Including the Gaussian feature
the chi-squared of the fit decreased by $\Delta\chi^2=-181.2$. To
check the significance of this excess, we applied the same method of
posterior predictive $p$-values described in the previous
section. This yielded a probability of $p\le 10^{-3}$ of the observed
improvement being due to chance. The best fitting values of the
parameters are listed in the column labelled as Model A (RGS) of
Table~\ref{tab1}.  A possible correlation with the composition of the
medium yielding interstellar absorption could be excluded, as even
varying the abundances of the atomic species with edges falling in the
considered energy band, a chi-squared similar to that obtained with
the Gaussian feature could not be obtained; at the same time,
contribution from absorption of \mbox{O\,{\sc viii}}, could be
excluded as the fit with an edge fixed at 0.87 keV did not determine a
fit of a comparable quality.

\subsection{Simultaneous RXTE - XMM-Newton spectrum }
\label{sec:xmmxte}

Analysis of the observation performed by {\xte} simultaneously to
{\xmm} allows to constrain on a broader energy band the X-ray spectrum
of {\src}. By using the same bands defined by \citet{watts2009} we
obtained values of the soft and hard colour of 1.1 and 1.0,
respectively. Comparing these values with those plotted in Fig.~4 of
\citet{watts2009}, we conclude that during this observation {\src} was
in the softer end of the hard island state.

\begin{figure}
\includegraphics[angle=0,width=\columnwidth]{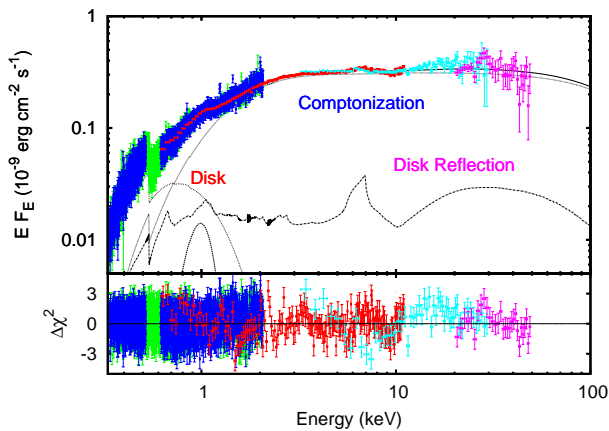}
\caption{0.33--50 keV unfolded spectrum of {\src} obtained fitting
  simultaneously spectra from RGS1/2 (0.33--2.07 keV, blue and green
  points, respectively), EPIC-pn (0.6--11 keV, red), PCA (3--30 keV,
  cyan) and HEXTE (20--50 keV, magenta) and considering Model B (see
  Table~\ref{tab1}), whose component are also over-plotted (top
  panel). Residuals with respect to Model B (bottom panel).}
\label{fig:spall}
\end{figure}

We fitted the spectra observed by PCA (3--30 keV) and HEXTE (20--50
keV) on-board {\xte} simultaneously to those obtained by RGS1/2
(0.33-2.07 keV) and by EPIC-pn (0.6--11 keV), fixing the normalisation
constant of the EPIC-pn spectrum to one, and letting a relative
normalisation constant free to vary for the other spectra. We obtained
a reduced chi-squared of 1.24 over 4458 degrees of freedom by using
Model B (see Table~\ref{tab1}), and letting the abundance of O in the
interstellar medium as a variable parameter.  Unlike the case
described in Section~\ref{sec:pn}, we significantly detected a
\texttt{diskbb} component even with a reflection component included in
the model, as its addition decreased the chi-squared by
$\Delta\chi^2=-432$ for two d.o.f. more. The Gaussian-shaped feature
at $\sim 1$ keV was described by parameters compatible with those
found when modelling both EPIC-pn and RGS data with model A. The
detection of a disc reflection component was significant also in the
broad-band spectrum, since removing it from the model yielded a
chi-squared increase of $\Delta\chi^2=314$ for five d.o.f. more. The
parameters of disc reflection are compatible with those indicated by
the analysis of EPIC-pn data alone, indicating an inclination of
$i=(30.6^{+3.7}_{-2.6})^{\circ}$ and an inner disk radius of $R_{\rm
  in}=25_{-11}^{+16}$ R$_g$. The best-fitting parameters are shown in
the rightmost columns Table~\ref{tab1}, the unfolded spectrum is
plotted in the upper panel of Fig.~\ref{fig:spall}, and residuals with
respect to the best-fit model shown in the lower panel.

To check if a slightly mis-calibrated response of the EPIC-pn at low
energies could affect model parameters, and in particular those
describing the reflection component, we also performed spectral
fitting after removing EPIC-pn data below 2 keV. The best-fitting
model had a reduced chi-squared of 1.22 (over 4427 d.o.f.) and no
significant difference with respect to the parameters listed in the
rightmost column of Table~\ref{tab1} was found. It is worth
  noting that in this case, the excess at $\approx 1$ keV was
  satisfactorily modelled by a reflection model with a Fe abundance of
  $2.0_{-0.20}^{+0.06}$ times the Solar value, without the need of an
  additional Gaussian feature. At the same time, given the slight
mismatch between the PCA and the EPIC-pn residuals (see bottom panel
of Fig.~\ref{fig:spall}), we removed the PCA data below 8 keV, but
still no significant variations were found in the model parameters.

\section{Temporal analysis}

In order to search the EPIC pn time series for a periodicity at the
377.3 Hz spin frequency of {\src}, we first reported the photon
arrival times to the Solar system barycentre using the source
coordinates of the optical counterpart estimated by
\citet{fox2005}. Since the pulse amplitude of AMSPs usually decreases
below 2 keV \citep{patruno2009,papitto2010}, compatible with the
increasing contribution of the accretion disc to the overall flux, we
restricted our analysis to the 2--11 keV energy band. We corrected
time delays induced by the orbital motion of the source in the binary
system using ephemeris determined by \citet{patruno2012}, who derived
a timing solution valid over a time interval of 2.6 yr (MJD
53539--54499) based on {\xte}/PCA data. To ascertain whether the
uncertainties on this set of parameters are small enough not to affect
coherence of the signal over the EPIC pn exposure, we considered the
relations given by \citet{caliandro2012}, who estimated the reduction
of the power obtained with a Fourier transform of a coherent signal,
$\epsilon^2$, as a function of the difference between actual orbital
parameters and the estimates used to correct photon arrival times. We
evaluated the relations they derived (see their Eq.~52, 64, 68, 72)
for the orbital and spin parameters of {\src} and a signal power
reduction of $\epsilon^2=0.8$; to produce a similar power decrease,
the estimates of semi-major axis of the NS orbit, eccentricity, epoch
of zero mean anomaly, and orbital period used in arrival time
corrections should differ by $\delta x=1.7\times10^{-3}$ lt-s, $\delta
e=0.04$, $\delta T^*=30$ s, $\delta P_{\rm orb}=2.7$ s, from the
actual values, respectively.  All but $\delta T^*$ are more than two
orders of magnitude larger than uncertainties quoted by both
\citet{patruno2012} and \citet{kaaret2006}, indicating how available
estimates were accurate enough not to induce loss of signal power. The
uncertainty on the epoch of zero mean anomaly was evaluated by
propagating the uncertainty on the estimate evaluated at the reference
epoch of the timing solution given by \citet{patruno2012}
($t_{0}=\rm{MJD}\: 53538.76$, Patruno, priv.~comm.), through the
$\Delta t=6.25$ yr elapsed until the {\xmm} observation \citep[see,
  e.g.][]{papitto2005}
\begin{eqnarray}
\sigma_{T^*}(t_{\rm EPN})=&\bigg\{[\sigma_{T^*}(t_{0})]^2+\left[\sigma_{P_{orb}}\frac{\Delta t}{P_{\rm orb}}\right]^2+\nonumber \\& \left[\frac{1}{2}P_{\rm orb}\dot{P}_{orb}\left(\frac{\Delta t}{P_{\rm orb}}\right)^2\right]^2 \bigg\}^{1/2} = 464 s,
\end{eqnarray}
 where $P_{\rm orb}=4995.2630(5)$ s is the orbital period,
 $\sigma_{T^*}=0.8$ s and $\sigma_{P_{orb}}=5\times10^{-4}$ s are the
 uncertainties on the epoch of zero mean anomaly and orbital period,
 respectively, and $\dot{P}_{orb}<1.2\times10^{-10}$ is the upper
 limit on the orbital period derivative at 95 per cent confidence
 level \citep{patruno2012}. As long as such loose upper limit on
 $\dot{P}_{orb}$ is considered, $\sigma_{T^*}(t_{\rm EPN})\simeq 15\:
 \delta T^*$, and $n_{T^*}=90$ corrections with values of $T^*$
 differing by $\delta T^*$, should have been performed to cover all
 the possible value of $T^*$, ranging from $T^{*}(t_{\rm
   EPN})-3\sigma_{T^*}(t_{\rm EPN})$ to $T^*(t_{\rm
   EPN})+3\sigma_{T^*}(t_{\rm EPN})$. However, the orbital period
 derivative measured from SAX J1808.4--3658 \citep{hartman2008,
   disalvo2008, hartman2009, burderi2009, patruno2012b}, a source with
 orbital parameters similar to those of {\src}, is $\sim 25$ times
 smaller than the upper limit given by Patruno (2012) for {\src}. The
 value measured for SAX J1808.4--3658 is already so large that largely
 non conservative scenarios of mass transfer should be invoked to
 explain it \citep{disalvo2008,burderi2009}. Consider an orbital
 period derivative equal to the value measured for SAX J1808.4--3658,
 we obtained $\sigma_{T^*}(t_{\rm EPN})=23.9$ s; only
 $n'_{T*}=6\:\sigma_{T^*}(t_{\rm EPN})/\delta T^*\simeq5$ trials were
 then needed to cover possible values of $T^*(t_{\rm EPN})$, when a
 plausible value of the orbital period derivative of {\src} was
 considered.

To determine over which frequency range a meaningful search for a
signal at the spin period of {\src} should have been performed, we
evaluated the spin frequency variation $\Delta\nu$ driven by the
accretion between 2005 and 2012. Assuming the base level spin
frequency derivative estimated by Patruno (2012),
$\dot{\nu}=4.2(1)\times10^{-14}$ Hz s$^{-1}$, compatible with the
torque expected to be imparted by mass accretion at a rate of few
$\times 10^{-10}$ M$_{\odot}$ yr$^{-1}$ onto a NS with magnetic field
$\simeq 10^{8}$ G, we obtained $\Delta\nu=8.3(2)\times10^{-6}$
Hz. Since the frequency Fourier resolution of a power spectrum
produced over the 60 ks EPIC pn time series is $\delta\nu_F=1/T_{\rm
  obs}=1.67\times10^{-5}$ Hz, one frequency bin contains both the
value of the spin frequency in 2005 ($377.29617188$ Hz, Patruno,
priv.~comm.) and the one expected in 2012 ($377.2961802(2)$ Hz). We
did not detect any significant signal in power spectra of the time
series corrected with the $n'_{T^*}=5$ plausible values of $T^*$,
inspecting a single frequency bin centred at 377.2961735 Hz, of width
equal to $\delta\nu_F$.  The maximum Leahy power we obtained is
$P_{\rm max}=4.7$, lower than the 3-$\sigma$ confidence level
threshold \citep{vanderklis1989,vaughan1994}
\begin{equation} 
  P_{3\sigma}(n_{tot}=5)=2\ln{(n_{tot})+2\ln{(p^{-1})}}=15.0.
\end{equation} 
This threshold is obtained by evaluating the power level that has a
probability lower than $p=2.7\times10^{-3}$ of being exceeded by noise
powers distributed as a $\chi^2$ with two degrees of freedom, taking
into account the five trials performed ($n'_{T^*}=5$, $n_{\nu}=1$). We
took into account the effect of noise-signal interactions due to the
fact that the total power in a frequency bin is the square of the sum
of noise and signal Fourier amplitude, each of which is a complex
number \citep{groth1975}. To this aim we used the routine given by
\citealt{vaughan1994}; the maximum observed power translates into an
upper limit on the signal power of $P_{\rm UL}=23.2$, at $3\sigma$
confidence level.  Considering that the total number of photons in the
time series is $N_{\gamma}=3.96\times10^6$, and that the average power
reduction due to frequency and $T^*$ binning is $\xi_{\nu}=0.773$
\citep{vaughan1994} and $\xi_{T^*}\simeq1-\epsilon^2/2=0.9$,
respectively, the 3-$\sigma$ upper limit on the pulse amplitude is
\citep{vaughan1994}
\begin{equation}
A_{UL}=\left\{\frac{2}{\xi_{\nu}\xi_{T^*}}\frac{P_{\rm UL}}{N_{\gamma}} \left[\mbox{sinc}\left({\frac{\pi}{2}\frac{\nu}{\nu_{Ny}}}\right)\right]^{-1}\right\}^{1/2}=4\times10^{-3}.
\end{equation}
Here, $\mbox{sinc}(x)=\sin(x)/x$, and the Nyquist frequency is
$\nu_{\rm Ny}=1/(2\:n_{\rm bin}\:t_{res})=2117$ Hz, as the time
resolution of the series is $t_{res}=2.952\times10^{-5}$ s and we set
$n_{\rm bin}=16$.

Even by extending the parameter space over which we searched for a
signal, inspecting all the values of T$^*$ compatible with the loose
available upper limit on the orbital period derivative ($n'_{T^*}=90$,
$n_{\nu}=1$, $P_{3\sigma}(n_{tot}=90)=20.8$), and the two nearest
Fourier independent frequencies ($n'_{T^*}=90$, $n_{\nu}=3$,
$P_{3\sigma}(n_{tot}=270)=23.0$), no detection was achieved ($P_{\rm
  max}=11.0$).  As some of the detections claimed by
\citet{patruno2012} were achieved on short time intervals, we
performed a pulsation search over $n_{s}=120$ data segments as short
as 500 s ($n'_{T^*}=5$, $n_{\nu}=1$, $P_{3\sigma}(n_{tot}=600)=24.6$),
but still found none significant at a 3$\sigma$ confidence level
($P_{\rm max}=19.7$).

\begin{figure}
\includegraphics[angle=0,width=\columnwidth]{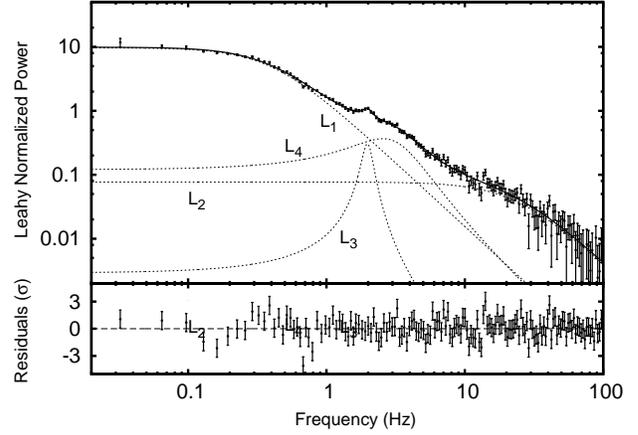}
\caption{\label{fig:pds} Power spectrum obtained by averaging 1940
  intervals, each $\simeq$31 s long, and re-binning the resulting
  spectrum with a factor 1.03. A white noise level of 1.992(2) was
  subtracted. The solid line is the best-fit model composed of four
  Lorentzian components, which are also shown as dotted lines (top
  panel). Residuals in units of $\sigma$ with respect to the best-fit
  model (bottom panel).}
\end{figure}

To study the aperiodic properties of the source, we averaged 1940
power density spectra produced over $\simeq$31 s long intervals,
retaining photons in the full 0.3--11 keV bandwidth, and re-binning
the resulting spectrum by a factor 1.03. Four Lorentzians components
were used to model the power spectrum obtained after the subtraction
of a white noise level of 1.992(2) Hz$^{-1}$ (see
Fig.~$\ref{fig:pds}$): two flat-top noise components (i.e. centred at
zero frequency), with width $W_1=0.80(1)$ and $W_2=47(7)$ Hz, and two
narrower Lorentzians centred at $\nu_3=1.98(2)$ and $\nu_4=2.6(2)$ Hz,
with width $W_3=0.4(1)$ and $W_4=3.6(3)$ Hz, labelled as $L_1$, $L_2$,
$L_3$ and $L_4$, respectively.  A similar decomposition is remarkably
similar to that shown by atoll sources in the hard state
\citep{olive1998}, and other AMSPs \citep{vanstraaten2005}. In
particular, the frequency and quality factor (defined as $\nu/W$) of
features we denoted as $L_3$ and $L_4$ are similar to those of the
quasi periodical oscillations identified by \citet{vanstraaten2005} as
$L_{\rm LF}$ and $L_h$, in low luminosity, non-pulsing atolls, such as
1E 1724--3045 and GS 1826--24. High frequency quasi periodical
oscillations were also searched for, by averaging 16 s long intervals,
but none was found.

\section{X-ray burst}

\begin{figure}
\includegraphics[angle=0,width=\columnwidth]{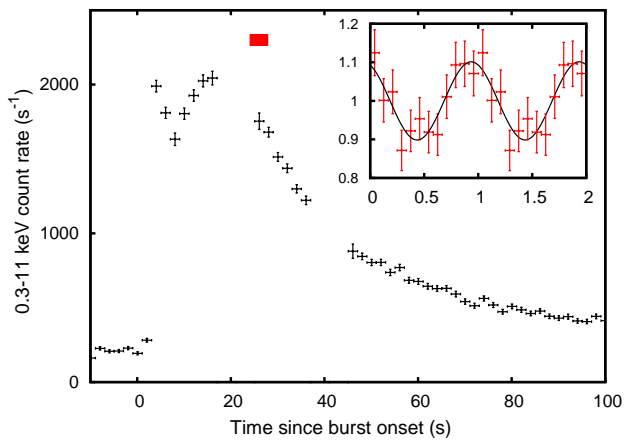}
\caption{\label{fig:bursttiming} 0.6--11 keV light curve of the burst
  observed by EPIC-pn on-board {\xmm}. Time is shown in seconds since
  the burst onset. The horizontal bar indicates the time interval
  during which a signal is barely detected at a Leahy power of 21.6.
  The inset shows the profile obtained by folding data obtained during
  such interval, around the  frequency, 377.04(25) Hz.}
\end{figure}

A type-I X-ray burst was observed during {\xmm} exposure (see inset of
Fig.~\ref{fig:lc} for the 0.6--11 keV light curve), starting on MJD
55823.69344. During first $\sim10$ s since the burst onset, EPIC-pn
count-rate saturated at a level of $\sim$ 1900 s$^{-1}$ due to
telemetry limitation, and the light curve cannot be considered
representative of the true burst light curve shape (see top panel of
Fig.~\ref{fig:bursttiming}). Presumably for the same reason, data were
not acquired during intervals 14 -- 24 and 34 -- 44 s, since the burst
start. At later times since the burst onset, the profile could be
modelled with an exponential decay with a time constant of
$\tau=46(1)$ s. This is significantly longer than the length of the
majority of bursts observed by {\xte} and {\it HETE}
\citep{suzuki2007,galloway2008,watts2009}, except for the burst
labelled as B1 in Fig.~2 of \citealt{watts2009}, and which also took
place where the source was in the hard state. However, there does not
seem to be a clear correlation of burst duration with the position of
the source in a colour-colour diagram; the position of the source
during the observation considered here is in fact much closer to that
shown by the source before shorter bursts (e.g. burst B4 in Fig.~4 of
\citealt{watts2009}), than at the onset of the other long burst
detected.

We searched for burst oscillations over 4 s intervals of EPIC-pn data;
we inspected frequencies ranging from 374 to 378 Hz, for a total of 16
trials in every interval, in order to cover the interval over which
the frequency of burst oscillations discovered by \citet{watts2009}
was observed to vary. We found a barely significant signal at a
frequency of 377.04(25) Hz with a Leahy power of 21.6, in the interval
starting 24 s after the burst onset. Such a power is larger than the
3$\sigma$ threshold for a single interval (17.2), but reduces to a
significance of 2.1$\sigma$ if the number of intervals considered to
analyse the burst (100) is taken into account; we cannot therefore
exclude it represents a statistical fluctuation. The folded profile is
shown in the inset of Fig.~\ref{fig:bursttiming}. If the
  oscillations were effectively present, this would be the first time
  they were observed in the hard state, since oscillations reported by
  \citet{watts2009} were found in a short duration burst, ignited when
  the source was in a soft state. We note that before the interval
where the marginal detection was achieved, sensitivity to pulsed
signals was severely reduced due to telemetry overload.

For the same reason, time-resolved spectroscopy of the burst could be
meaningfully performed only after $\sim 25$ s since the burst
onset. No background was subtracted, and the best-fit Model A (see
leftmost column of Table~\ref{tab1}) was used to model persistent
emission, letting only the absorption column free to vary. Spectra in
the 0.6--11 keV energy band were satisfactorily fit by an absorbed
black-body, with a temperature decreasing from 1.6(1) to 0.9(1) keV at
the end of the burst, and an apparent radius roughly constant around a
value of $\sim 6$ d$_5$ km.  The value of the absorption column
density we found, $N_{\rm H}=(0.99\pm0.15)\times10^{21}$ cm$^{-2}$ is
compatible with estimates indicated by analysis of {\it persistent}
emission (see Table~\ref{tab1}).

\section{Discussion and conclusion}

LMXB hosting a NS are usually classified in Z and atoll sources, based
on the track followed in a colour-colour diagram and the aperiodic
time variability \citep{hasinger1989}, with atolls emitting a
luminosity $\simlt 0.5$ L$_{Edd}$, and showing a harder spectrum, on
average. Similar to other AMSP, {\src} behaves as a typical atoll
source, spending most of the time in the hard island state
\citep{watts2009}. Also the observations considered here caught the
source in such state, while emitting a 0.33--50 keV unabsorbed
luminosity of $4.07(3)\times10^{36}$ erg s$^{-1}$; roughly $90$ per
cent of the X-ray observed in such band is described by a
$\Gamma\simeq 2$ hard component, with no cut-off significantly
detected in our data. To physically model such a component we used
Comptonization of soft, $kT_{\rm soft}\simeq 0.45$ keV, photons in a
medium with an electron temperature fixed to 100 keV. The slope of
such hard component and the assumed electron temperature indicate an
optical depth, $\tau\simeq 1$ \citep{zdziarski1996,zycki1999}. A
similar Comptonized component was used by \citet{falanga2007} to model
the 2--300 keV spectrum observed by {\xte} and {\it INTEGRAL} during
observations performed in 2005, and is invariably found to dominate
the X-ray spectrum of AMSP \citep{poutanen2006}. Considering also the
energy-dependent properties of pulse profiles, \citet{gierlinski2002}
interpreted this component as originated in the accretion columns,
where mass is channelled by the magnetic field and heated by the
interaction with radiation coming from the NS surface. However, the
non detection of pulsations casts doubts on the presence of
accretion columns during the observation of {\src} presented here.

Two broad emission features centred at energies 0.98(1) and 6.58(7)
keV were detected in EPIC-pn spectrum. The energy of the latter is
compatible with K$\alpha$ transition of \mbox{Fe\,{\sc xxiii}--{\sc
    xxv}} \citep[6.6--6.7 keV, see, e.g.,][]{kallman2004}; such
detection confirms the results previously obtained by
\citet{cackett2010} with a {\it Suzaku} observation, and by
\citet{falanga2007}, even if at a much lower spectral
resolution. Origin of the $\approx 1$ keV feature is instead less
clear. Calibration uncertainties of the EPIC-pn spectra taken in
timing mode were often invoked to explain appearance of strong
features at energies $\simlt 1$ keV; however, the presence of a
similar line-shaped residual also in spectra observed by EPIC-MOS2 and
RGS strongly suggests a physical origin.  Both the features have a
broadness $\sigma/E\simeq0.1$, compatible with a similar mechanism
determining their shape. The easiest interpretation of such width is
in terms of reflection of hard X-ray photons in the accretion disc,
where the strong velocity field broadens discrete features, and
relativistic effects distort and smear their shape, shifting them
towards lower energies.  In the case of the observations analysed
here, a mildly ionised $\xi\simeq750$ erg cm s$^{-1}$ disc reflection
component, broadened by motion in an accretion disc and representing
$10$ per cent of the overall flux, successfully models the iron line
and improves description of the spectral continuum. However, a
self-consistent modelling of the two discrete features, could not be
obtained, as far as the EPIC-pn data below 2 keV are considered.
Strong emission features at energies similar than those found here
were recently reported thanks to {\xmm} spectra of a couple of
narrow-line Seyfert 1 galaxies, and interpreted in terms of emission
lines due to Fe K$\alpha$ and Fe L$\alpha$ transitions taking place as
the accretion disc reflects hard illuminating emission
\citep{fabian2009,ponti2010,zoghbi2010,fabian2012}. Despite the ratio
between the photon fluxes in the two features observed by EPIC-pn
($=6.8 \pm 1.7$) is less extreme than those they reported
($\simgt20$), it is still too large to be described by a single
reflection environment like that we considered. A lower flux ratio is
in fact expected when the illuminating spectrum becomes softer than
$\Gamma\simlt2$ at a given flux, and iron is less ionised
\citep{ross2005}.  However, excluding from the fit EPIC-pn data below
2 keV and modelling low energies with RGS data alone, we could find a
self-consistent modelling of the two features in the same reflecting
environment, with an iron abundance $\approx2$ times larger than the
Solar value.  Another, not necessarily alternative interpretation of
the excess is in terms of K$\alpha$ transition of overabundant
helium-like Ne; though, it was not possible to vary the abundance of
this element in the reflection model, and simply adding a blurred
feature at the relevant energy did not result in a completely
satisfactory description. The failure in deriving a simultaneous
description of the feature detected by EPIC-pn possibly follows from
superimposition of different transitions not completely taken into
account by the reflection model used, different abundances of ionised
elements, or because of residual calibration uncertainties. The
satisfactory modelling of the two features by a reflection component
with overabundant iron, obtained when EPIC-pn data at low energies are
discarded, favours the latter hypothesis. In any case, the
simultaneous detection of such feature from different instruments
makes it a very promising additional probe of the accretion flow to
LMXB hosting NS, the nature of which will be firmly assessed by future
observations.

Fitting the smeared reflected component with the \texttt{rdblur}
kernel of \citet{fabian1989}, we estimated the inclination, the
emissivity, the inner radius of the disc,
$i=(30.6^{+3.7}_{-2.6})^{\circ}$, $\beta\simeq -3.7^{+1.1}_{-2.5}$,
$R_{\rm in}\simeq 25^{+16}_{-11}\:R_g$, respectively. From Doppler
tomography of a broad H$\alpha$ line assumed to originate in the
accretion disc, \citet{elebert2008} estimated the inclination of the
system to be $i\simlt20^{\circ}$ for a 1.4 $M_{\odot}$, a value not
far from that reported here.  Our estimate of the inner disc radius is
compatible with the value of $14(1)$ R$_g$ found by \citet[][for a
  rather low inclination of 4$^{\circ}$]{cackett2010}, who modelled a
{\it Suzaku} spectrum of this source taken when its 0.5--25 keV
luminosity was $\simeq 5\times10^{35}$ erg s$^{-1}$ (roughly a factor
three times lower than during the observations presented here) by
adding a reflection component to a $\Gamma\simeq2.2$ power law. At a
distance of $\approx 25\:R_g$ from the compact object, relativistic
effects are small and the line shape we observed is rather
symmetric. However, alternative interpretations proposed to explain
broad symmetric iron lines, such as Compton broadening in a hot
corona, are not compatible with what observed in the broadband
spectral decomposition. An optically thin, $\tau\sim 1$ corona causes
an average photon energy shift of
$\Delta\epsilon/\epsilon=(4kT_e-\epsilon)/m_ec^2$ \citep{
  rybicki1979}; to produce the observed line width of $0.7(1)$ keV,
the electrons in the Comptonizing cloud should have a temperature of
$\simeq15$ keV and a similar component was not observed in the
spectrum. The value estimated for the inner disc radius translates
into a value of 52$_{-23}^{+33}$ km, for a 1.4 $M_{\odot}$ NS. Despite
the large uncertainty affecting this estimate ($R_{\rm in}>12$ km at
99 per cent confidence level), the shape of reflected spectrum 
indicates that the accretion disc is truncated. A similar size of the
inner rim of the disc is suggested also by the $kT_{\rm in}\simeq0.2$
keV thermal component detected at low energies; if modelled with an
accretion disc model, its normalisation indicates an apparent inner
disc radius $\simeq 30$ d$_5$ km (for $i=30^{\circ}$). The actual
inner radius of the disc can be larger up to a factor of two than this
estimate, when colour corrections are considered \citep{merloni2000}
and torque is assumed to be retained through the transition region
from the optically thick disc to the inner hot flow
\citep{gierlinski1999}. At the same time, considering the luminosity
of the photo-ionising Comptonized component ($L_{\rm irr}\simeq
3.5\times10^{36}$ d$_5^2$ erg s$^{-1}$), the ionisation parameter of
the observed reflected component (${\xi}=L_{\rm irr}/n_{\rm H} \rho^2
\simeq 750$ erg cm s$^{-1}$) is expected to be produced at a distance
$\rho\simeq 30$ km from the source of illuminating photons, for a
hydrogen density of $5\times10^{20}$ cm$^{-3}$, a value appropriate
for a Shakura Sunyaev disc around an accreting NS \citep[see, e.g.,
][]{ross2007}. Further, for a 10 km NS and a system inclination of
$30^{\circ}$, the observed ratio between the reflected and
illuminating component ($\simeq 0.1$, corresponding to a reflection
amplitude $\Omega/2\pi\simeq0.13$) indicates an inner disc radius
$\simgt 50$ km \citep[][who calculated the reflection amplitude as a
  function of the inner disc radius in the case of a small
  illuminating spot at the rotational pole of the NS; a more extended
  source of illuminating photons produces a larger reflection
  amplitude at a fixed disc radius]{gierlinski2005}.

The main peculiarity of {\src} is coherent pulsations intermittence,
whose causes are not yet fully understood. \citet{cumming2008}
proposed that intermittence is caused by a screening of the
magnetosphere and burial of the magnetic field under the NS
surface. On the other hand, \citet{romanova2008} and \citet{lamb2009},
while involving the presence of a magnetosphere, explained
intermittence in terms of penetration of in-falling matter through the
magnetic field lines, and of movements of a hot spot located close to
the spin axis of the NS, respectively, depending on variations of the
mass accretion rate.  Alternative scenarios proposed to explain the
paucity of pulsars among LMXB, and possibly related to pulse
intermittence, include scattering in a hot, relatively optically thick
cloud surrounding the NS \citep{titarchuk2002}, and reduction of pulse
amplitude due to gravitational light bending
\citep{ozel2009}. However, considering how a Comptonized component
produced in an optically thin, $\tau\approx1$, corona was observed
regardless of pulse detection, and how the compactness of the NS
hardly varies on timescales of few years, these two latter
interpretations do not reproduce easily the behaviour observed from
{\src}. The evidences we obtained from spectral analysis of an
accretion disc truncated quite far from the NS have to be discussed in
view of the time variability shown by the source. No pulsations were
detected during the persistent emission, with a very low upper limit
of 0.4 per cent, at 3 $\sigma$ confidence level, on the 2--11 keV
pulsed fractional amplitude. Such a limit is of the same order of the
weaker pulsations detected by \citet{patruno2012}. As already
suggested by \citet{galloway2008}, this source could persistently show
pulsations at a very low, $\simlt 0.1$ per cent, amplitude, beyond the
detection sensitivity of current instruments. The inner disc radius
derived by spectral modelling of the reflection component is
compatible with models in which a magnetosphere is able to truncate
the optically thick disc, and coherent oscillations are reduced in
amplitude by geometrical effects \citep{lamb2009} and/or by weak
channelling of accreting matter, if any \citep{romanova2008}. However,
the tight upper limit on pulse amplitude and the absence of a clear
relation between appearance of pulses and variations of the mass
accretion rate, do not necessarily favour them. On the other hand,
spectral and aperiodic timing properties of this source are entirely
indistinguishable to that of (assumed) non pulsing atolls in the hard
state, suggesting how the accretion flow to {\src} is probably not
channelled by a magnetosphere. This would keep open the possibility of
a buried magnetic field \citep{cumming2008}. In such a scenario, the
optically thick flow in a geometrically thin accretion disc would be
replaced by a inner optically thin hot flow, rather than by a
magnetospheric flow. This fits into the truncated disc scenario
explaining hard states shown by both accreting black holes and neutron
stars \citep[see, e.g.,][for a review]{done2007}. Similar estimates of
the inner disc radius were indeed obtained from iron line modelling of
non-pulsing atolls in the hard state \citep[e.g., the cases of 4U
  1705--44 and MXB 1728-34;][]{dai2010,egron2011,egron2012}. It
remains an open question whether the accretion disc flow is truncated
by a similar mechanism, connected to the overall evolution of the
accretion disc rather than to the NS magnetic field, even when a
magnetosphere is certainly present (i.e., when pulsations are
detected). Long observations of AMSPs performed while they show
pulsations, and aimed at a detailed modelling of broad emission
features, will be crucial to this end.

During the {\xmm} observation a type-I X-ray burst was
detected. Because of telemetry overflow, we could not determine the
peak burst luminosity, nor if a photospheric radius expansion took
place. We barely detected a signal at a frequency of 377.04(25) Hz
during a 4 s interval, 24 s after the burst onset; however, the
observed power is low, and the signal is significant only if the
number of burst intervals inspected is not taken into account.  Burst
oscillations from this source were already found by \citet{watts2009}
during a short burst ($\tau=7$--$8$ s). The burst during which they
detected pulsations ignited when the source was in the soft (banana)
state, with a a 2--16 keV flux of $\approx63$ mCrab as observed by
{\xte}/PCA, which is to date the highest persistent flux at which
bursts have been detected in this source. Oscillations were detected
during the initial stages of the burst decay, at a frequency $\sim 1$
Hz below the spin frequency of the source, drifting towards this value
as the burst evolved.  On the other hand, the 46 s time decay of the
burst reported here, the persistent source flux (24 mCrab in the 2--16
keV range) and the hard spectral state are not easily reconciled with
the general picture of the non-pulsing LMXBs and intermittent pulsars,
which preferably show oscillations during short bursts ignited when
the source is in the soft state, presumably accreting at a rate larger
than average \citep{galloway2008}. Moreover, the initial burst
oscillation frequency would be within few tenths of Hz from the spin
frequency of the source, 377.296 Hz, as it is commonly observed for
oscillations occurring during the decay of bursts shown by
persistently pulsing AMSP \citep{watts2012}; though, as the
oscillations barely detected here were observed at later times since
the burst onset, such frequency could be compatible with the positive
frequency drift observed by \citet{watts2009}. Assuming burst
oscillations marginally detected during this observation to be real,
they would be much more similar to those of persistently pulsing AMSP,
in terms of spectral state at the burst ignition and frequency drift,
than those detected by \citet{watts2009}; this could indicate a role
played by a residual magnetic field during the burst reported here,
under the hypothesis that the still poorly understood properties of
burst oscillations are related to the channelling of accreted material
by the NS magnetic field \citep[see, e.g., the review
  by][]{watts2012}.

\section*{Acknowledgments}

This work is based on observations obtained with {\xmm}, an ESA
science mission with instruments and contributions directly funded by
ESA Member States and NASA. We thank Evan Smith and the whole {\xte}
team for re-scheduling an observation already planned (PI:
D.~Galloway), to provide a simultaneous observation to the {\xmm}
pointing.  AP acknowledges the support of the grants AYA2012-39303 and
SGR2009-811, as well as of the iLINK program 2011-0303. LB, TDS and EE
acknowledge the support of Initial Training Network ITN 215212,
``Black Hole Universe'', funded by the European Union. We thank
Alessandro Patruno for useful discussions and the reviewer for
constructive comments and suggestions.

\bibliography{../../biblio}
\bibliographystyle{mn2e}

\end{document}